\documentclass[fleqn]{revtex4}

\usepackage{amsmath}
\usepackage{graphicx}
\usepackage{dcolumn}

\begin{document}

\title{Bethe logarithm for the helium atom}

\author{Vladimir I. Korobov}
\affiliation{Bogoliubov Laboratory of Theoretical Physics, Joint Institute for Nuclear Research, Dubna 141980, Russia\\
Peoples' Friendship University of Russia (RUDN University), 6 Miklukho-Maklaya St, Moscow, 117198, Russia}

\begin{abstract}
The Bethe logarithm for a large set of states of the helium atom is calculated with a precision of 12-14 significant digits. The numerical data is obtained for the case of infinite mass of a nucleus. Then we study the mass dependence and provide coefficients of the $m_e/M$ expansion, which allows us to calculate accurate values for the Bethe logarithm for any finite mass. An asymptotic expansion for the Rydberg states is analyzed and a high quality numerical approximation is found, which ensures 7-8 digit accuracy for the $S$, $P$, and $D$ states of the helium atom.
\end{abstract}

\maketitle

\section{Introduction}

The Bethe logarithm is an important constituent of the leading order radiative contribution to the energy of light atoms \cite{BS}. In the case of the hydrogen atom the splitting between $2s$ and $2p_{1/2}$ levels of about 1058 MHz detected in the Lamb and Retherford experiment \cite{Lamb47}, also known as the Lamb shift, was explained by Hans Bethe \cite{Bethe47} in a purely nonrelativistic fashion by introducing the mean excitation energy $K_0$. However, already in the case of a helium atom it was realized that to obtain numerically the mean excitation energy $K_0$, or more precisely, the Bethe logarithm $\ln(K_0/R_\infty)$, is a very nontrivial task \cite{KabirSalpeter}. It is astonishing that one of the earliest calculations \cite{Schwartz61} remained the most accurate for more than 30 years!

The significant breakthrough in numerical studies of the Bethe logarithm had been achieved in 90s of last century. Almost simultaneously three independent groups \cite{Morgan,KorobovBL99,DrakeCJP} using different approaches reached a precision which exceeded the earlier results by three to five orders of magnitude! Since that time precision calculations of many other atomic and molecular systems appeared. Among them are calculations of ro-vibrational states in the hydrogen molecular ions H$_2^+$ and HD$^+$ \cite{Korobov12bl}, adiabatic two electron calculation of the Bethe logarithm for molecular hydrogen \cite{Pachucki09bl}, three electron Li-like \cite{Drake03bl,Pachucki03bl}, and four electron Be-like atoms and ions \cite{Pachucki04bl,Bubin10bl}.

In this work we present a comprehensive calculation of the Bethe logarithm for the helium atom. This study covers a wide range of orbital angular momentum states up to and including $F$ states. We provide necessary numerical tools, which allow to calculate a precise value for the Bethe logarithm for a helium atom with a finite nuclear mass $M$ as well as for high $n$ states, where $n$ is the principal quantum number of an excited electron.

Atomic units are used throughout.

\section{Numerical approach to calculate the Bethe logarithm}

In this section we follow the method described in detail in \cite{Kor12}. We define the Bethe logarithm as the ratio
\begin{equation}\label{bethe_ratio}
\beta(L,v) = \frac{\mathcal{N}}{\mathcal{D}}\>.
\end{equation}
where the numerator $\mathcal{N}(n,L)$ is expressed by the following integral:
\begin{subequations}\label{bethe}
\begin{equation}\label{numerator}
\mathcal{N}(n,L) =
   \int_0^{E_h}\,k\,dk
   \left\langle
      \mathbf{J}\left(\frac{1}{E_0\!-\!H\!-\!k}+\frac{1}{k}\right)\mathbf{J}
   \right\rangle
   +\int_{E_h}^\infty\,\frac{dk}{k}\,
   \left\langle
      \mathbf{J}\frac{(E_0\!-\!H)^2}{E_0\!-\!H\!-\!k}\mathbf{J}
   \right\rangle.
\end{equation}
while the denominator is
\begin{equation}\label{denominator}
\mathcal{D}(n,L) =
   \Bigl\langle
      \mathbf{J}\left[H,\mathbf{J}\right]
   \Bigr\rangle =
   \frac{\bigl\langle
      \left[\mathbf{J}\left[H,\mathbf{J}\right]\right]
   \bigr\rangle}{2}\>.
\end{equation}
\end{subequations}
Here $E_h$ is the Hartree energy (the atomic unit of energy), $\mathbf{J}$ is the nonrelativistic electric current density operator of the atomic system
\begin{equation}\label{current}
\mathbf{J}=\sum_i \frac{z_i}{m_i}\mathbf{P}_i,
\end{equation}
and $z_i$, $m_i$ are the charges and masses of the particles.

\subsection{First order perturbation wave function, $\psi_1(\cdot)$, and asymptotic expansion of $J(k)$ at $k\to\infty$.}

The key quantity for our numerical studies is
\begin{equation}\label{Jk}
J(k) =
   \left\langle
      \mathbf{J}\left(E_0\!-\!H\!-\!k\right)^{-1}\mathbf{J}
   \right\rangle.
\end{equation}
Knowing this function one immediately gets a value for the nonrelativistic Bethe logarithm using Eq.~(\ref{bethe}). Relation between $J(k)$ and other forms of the integrand in~(\ref{bethe}) may be found in \cite{Schwartz61}.

A general procedure to calculate $J(k)$ is to solve the equation
\begin{equation}\label{eq_psi1}
(E_0-H-k)\psi_1 = i\mathbf{J}\psi_0,
\end{equation}
for different values of $k$. Since we are interested in the asymptotic behaviour of $J(k)$ for $k\to\infty$, it is assumed that $k$ is sufficiently large and as a first approximation one may take
\begin{equation}
\psi_1^{(0)} = -(i/k)\mathbf{J}\psi_0.
\end{equation}

Any approximate solution for $\psi_1$ may be (formally) iteratively improved:
\begin{equation}
\begin{array}{@{}l}\displaystyle
\psi_1^{(n)} =
   -\frac{i}{k}\mathbf{J}\psi_0+\frac{1}{k}(E_0\!-\!H)\psi_1^{(n-1)}
\end{array}
\end{equation}
and the next iteration would be
\begin{equation}\label{psi1^1}
\psi_1^{(1)} =
   -\frac{i}{k}\mathbf{J}\psi_0+\frac{1}{k^2}\left[H,i\mathbf{J}\right]\psi_0
\end{equation}
where
\begin{equation}\label{HJ}
[H,i\mathbf{J}] =
   \sum_{i>j} z_iz_j\left(\frac{z_j}{m_j}-\frac{z_i}{m_i}\right)\frac{\mathbf{r}_{ij}}{r_{ij}^3}\>,
\qquad
\mathbf{r}_{ij}=\mathbf{r}_j\!-\!\mathbf{r}_i.
\end{equation}

At small $r_{ij}$, $\psi_1$ should be smooth. In order to get a proper solution, one has to consider Eq.~(\ref{eq_psi1}) for $r_{ij}\to0$ and keep only important terms (here $m_{ij} = m_im_j/(m_i+m_j)$ is the reduced mass of a pair, we also use a notation $(\cdot)$ in the list of variables of a function if the particular form of the internal variables is not important)
\[
\left(\frac{1}{2m_{ij}}\Delta_{ij}-k\right)\psi_1(r_{ij},\cdot) = 0
\]
that gives homogeneous solutions of the type
\[
\sim\,\frac{\mathbf{r}_{ij}}{r_{ij}^3}\>e^{-\mu_{ij}r_{ij}}\,(1+\mu_{ij}r_{ij})
\]
with $\mu_{ij}=\sqrt{2m_{ij}k}$. These solutions, taken for different pairs of particles, may be added to $\psi_1^{(1)}$ to make the whole wave function smooth. So, we come to an approximation of $\psi_1$ for $k\!\to\!\infty$, which has the following form,
\begin{equation}\label{eq:psi1}
\psi_1^{(1)} =
   -\frac{i}{k}\mathbf{J}\psi_0(\cdot)
   +\frac{1}{k^2}\sum_{i>j}
      z_iz_j\left(\frac{z_j}{m_j}-\frac{z_i}{m_i}\right)
         \frac{\mathbf{r}_{ij}}{r_{ij}^3}
      \left[
         1-e^{-\mu_{ij}r_{ij}}\,(1+\mu_{ij}r_{ij})
      \right]\psi_0(\cdot)\,.
\end{equation}
As is seen from this equation, there is no singular term in the wave function corresponding to a pair of identical particles.

The integrand $J(k)$ may be evaluated using the variational formalism as a stationary solution of a functional of $\psi_1$
\[
J(k) =
   -2\left\langle \psi_0 |i\mathbf{J}| \psi_1 \right\rangle
   -\left\langle
      \psi_1 (E_0\!-\!H\!-\!k) \psi_1
   \right\rangle.
\]
To derive the asymptotic expansion we substitute $\psi_1^{(1)}$ into the functional. That results in \cite{Kor12}:
\begin{equation}\label{expansion}
\begin{array}{@{}l}\displaystyle
J(k) =
   -\frac{1}{k}\left\langle\mathbf{J}^2\right\rangle
   -\frac{1}{k^2}\,\frac{
      \left\langle\>
         \left[i\mathbf{J}\!,\left[H,i\mathbf{J}\right]\right]
      \>\right\rangle}{2}
   -\frac{1}{k^3}
      \sum_{\genfrac{}{}{0pt}{}{i>j,k>l}{(i,j)\ne (k,l)}}
         z_iz_jz_kz_l
         \left(\frac{z_i}{m_i}-\frac{z_j}{m_j}\right)
         \left(\frac{z_k}{m_k}-\frac{z_l}{m_l}\right)
         \left\langle\frac{\mathbf{r}_{ij}\mathbf{r}_{kl}}{r_{ij}^2r_{kl}^2}\right\rangle
\\[4mm]\displaystyle\hspace{12mm}
   -\frac{1}{k^3}
   \sum_{i>j}z_i^2z_j^2
   \left(\frac{z_i}{m_i}-\frac{z_j}{m_j}\right)^2
   \biggl\{
   4\pi\mathcal{R}_{ij}+
   \Bigl[
      \sqrt{2m_{ij}k}
      +z_iz_jm_{ij}\Bigl(\ln(m_{ij}k)\!-\!\ln{2}\!-\!1\Bigr)
   \Bigr]
   4\pi\left\langle\delta(\mathbf{r}_{ij})\right\rangle
   \biggr\}+\dots
\end{array}
\end{equation}
Here $\mathcal{R}_{ij}$ is a finite functional, which replaces a divergent $1/r^4$ operator, and is defined by the following expression
\begin{equation}\label{R}
\mathcal{R}_{ij} =
   \lim_{\rho\to0}
   \left\{
   \left\langle
      \frac{1}{4\pi r_{ij}^4}
   \right\rangle_{\!\!\rho}\!
   -\left[
      \frac{1}{\rho}\left\langle\delta(\mathbf{r}_{ij})\right\rangle
      +\left(\ln{\rho}\!+\!\gamma_E\right)\left\langle\delta'(\mathbf{r}_{ij})\right\rangle
   \right]
   \right\}
\end{equation}
where
\[
\left\langle\phi_1|\delta'(\mathbf{r})|\phi_2\right\rangle =
\left\langle\phi_1\left|
   \frac{\mathbf{r}}{r}\boldsymbol{\nabla}\delta(\mathbf{r})
\right|\phi_2\right\rangle =
   -\left\langle\partial_r\phi_1|\delta(\mathbf{r})|\phi_2\right\rangle
   -\left\langle\phi_1|\delta(\mathbf{r})|\partial_r\phi_2\right\rangle.
\]

The mixed terms: $(\mathbf{r}_{ij}\mathbf{r}_{kl})/(r_{ij}r_{kl})^2$ are finite and do not require any regularization.

\subsection{Variational property}

If we consider the quantity
\begin{equation}
\mathcal{J}_\Lambda =
   \int_0^{\Lambda} k\,dk\>J(k) =
   \sum_n \bigl|\left\langle\psi_0|\mathbf{J}|\psi_n\right\rangle\bigr|^2
      \left[
         \Lambda-(E_0\!-\!E_n)\!\ln{\left|
                 \frac{E_0\!-\!E_n}{E_0\!-\!E_n\!-\!\Lambda}\right|}
      \right].
\end{equation}
we would find that for the ground state of a system this quantity possesses the variational property, since for the integrand for all $k$ the following inequality is fulfilled
\[
J_{\rm exact}(k) \ge J_{\rm numerical}(k).
\]
The same property remains satisfied for other states if integration is performed from some $k_0\sim 1$, which lies above the poles related to the states $E_n<E_0$. It is known from practical calculations that the low $k$ contribution becomes numerically converged to a high accuracy at a moderate basis length of intermediate states, and thus with a good confidence the variational property, the higher the value of $\mathcal{J}_\Lambda$ the more accurate solution, is still remained in force. That allows us to perform optimization of the variational parameters of the basis set.

\subsection{Numerical scheme}

Here we consider the numerical scheme for the three-body Coulomb problem, which is then used in calculations of the Bethe logarithm for the helium and $\mbox{H}_2^+$ ground states. The wave functions both for the initial bound state and for the first order perturbation solution (or intermediate state), are taken in the form,
\begin{equation}\label{var}
\Psi_L(l_1,l_2) = \sum_{i=1}^{\infty}
    \Big\{
       U_i\,{\rm Re}\bigl[e^{-\alpha_i r_1-\beta_i r_2-\gamma_i r}\bigr]
      +W_i\,{\rm Im}\bigl[e^{-\alpha_i r_1-\beta_i r_2-\gamma_i r}\bigr]
\Big\}\mathcal{Y}^{l_1l_2}_{LM}(\mathbf{r}_1,\mathbf{r}_2),
\end{equation}
where $\mathcal{Y}^{l_1l_2}_{LM}(\mathbf{r}_1,\mathbf{r}_2)$ are the solid bipolar harmonics as defined in \cite{Varshalovich}, $r_1$, $r_2$ and $r$ are the Hylleraas coordinates of the two electrons in the helium atom, $\mathbf{r}=\mathbf{r}_2-\mathbf{r}_1$, and $L$ is a total orbital angular momentum of a state. Complex parameters $\alpha_i$, $\beta_i$ and $\gamma_i$ are generated in a quasi-random manner
\cite{Kor00}:
\begin{equation}
\begin{array}{l}\displaystyle
\alpha_i =
   \left[\left\lfloor
            \frac{1}{2}i(i+1)\sqrt{p_\alpha}
         \right\rfloor(A_2-A_1)+A_1\right]
 +i\left[\left\lfloor\frac{1}{2}i(i+1)\sqrt{q_\alpha}
         \right\rfloor(A'_2-A'_1)+A'_1\right],
\end{array}
\end{equation}
$\lfloor x\rfloor$ designates the fractional part of $x$, $p_\alpha$ and $q_\alpha$ are some prime numbers, say, 2, 3, 5, etc., $[A_1,A_2]$ and $[A'_1,A'_2]$ are real variational intervals which need to be optimized. The parameters $\beta_i$ and $\gamma_i$ are obtained in a similar way.

The basis set for intermediate states is constructed as follows:
\begin{enumerate}
\item
First we use a regular basis set, which is taken similarly to the initial state with regular values of the parameters $(\alpha,\beta,\gamma)$ in the exponentials.
\item Then we build a special basis set with exponentially growing parameters for a particular $r_{ij}$
\begin{equation}\label{basis_2}
\left\{
\begin{array}{@{}ll}\displaystyle
A_1^{(0)} = A_1, & A_2^{(0)} = A_2
\\[1mm]\displaystyle
A_1^{(n)} = \tau^n A_1, \qquad & A_2^{(n)} = \tau^n A_2
\end{array}\right.
\end{equation}
where $\tau=A_2/A_1$.

\begin{table}[t]
\caption{The Bethe logarithm calculations for the ground and excited states of the helium atom with infinite nuclear mass $M_{\rm He}\to +\infty$ and comparison with most precise previous calculations.}\label{tb:results}
\begin{center}
\begin{tabular}{c@{\hspace{5mm}}l@{\hspace{5mm}}l@{\hspace{5mm}}l@{\hspace{5mm}}l@{\hspace{5mm}}l@{\hspace{5mm}}l}
\hline\hline
\vrule height12pt width0pt
$n$ & \multicolumn{1}{c}{$n^{\,1\!}S~$} & $~~~~~~~~n^{\,3\!}S$ & $~~~~~~~~n^{\,1\!}P$ & $~~~~~~~~n^{\,3\!}P$ \\
\hline
\vrule height11pt width0pt
 1 & 4.370\,160\,223\,0703(3)  \\
   & 4.370\,160\,218(3)$^a$   \\
   & 4.370\,160\,222\,9(1)$^b$ \\
   & 4.370\,160\,223\,06(2)$^c$ 
\\[1.5mm]
 2 & 4.366\,412\,726\,417(1)  & 4.364\,036\,820\,476(1)   & 4.370\,097\,743\,554(2)  & 4.369\,985\,364\,549(3) \\
   & 4.366\,412\,72(7)$^a$    & 4.364\,036\,82(1)$^a$     & 4.370\,097\,82(3)$^a$    & 4.369\,985\,20(2)$^a$   \\
   & 4.366\,412\,726\,2(1)$^b$& 4.364\,036\,820\,41(2)$^b$& 4.370\,097\,743\,5(1)$^b$& 4.369\,985\,364\,4(2)$^b$
\\[1.5mm]
 3 & 4.369\,164\,860\,824(2)  & 4.368\,666\,996\,159(2)   & 4.370\,295\,862\,299(4)  & 4.370\,235\,654\,775(4) \\
   & 4.369\,164\,871(8)$^a$   & 4.368\,666\,92(2)$^a$     & 4.370\,295\,75(9)$^a$    & 4.370\,233\,9(2)$^a$
\\[1.5mm]
 4 & 4.369\,890\,632\,356(3)  & 4.369\,723\,392\,715(4)   & 4.370\,363\,160\,331(5)  & 4.370\,334\,604\,477(5) \\
   & 4.369\,890\,66(1)$^a$    & 4.369\,723\,44(5)$^a$     & 4.370\,363\,2(2)$^a$     & 4.370\,334\,16(5)$^a$
\\[1.5mm]
 5 & 4.370\,151\,796\,310(4)  & 4.370\,078\,509\,668(4)   & 4.370\,390\,514\,367(5)  & 4.370\,375\,352\,464(5) \\
   & 4.370\,151\,7(1)$^a$     & 4.370\,078\,31(8)$^a$     & 4.370\,390\,54(4)$^a$    & 4.370\,374\,6(2)$^a$
\\[1.5mm]
 6 & 4.370\,266\,974\,319(5)  & 4.370\,229\,062\,747(5)   & 4.370\,403\,502\,993(6)  & 4.370\,394\,624\,37(2)
\\[1.5mm]
 7 & 4.370\,325\,261\,772(5) & 4.370\,303\,319\,792(5) \\
\hline
\vrule height11pt width0pt
 $n$ & $~~~~~~n^{\,1\!}D$ & $~~~~~~n^{\,3\!}D$ & $~~~~~~n^{\,1\!}F$ & $~~~~~~n^{\,3\!}F$ \\
\hline
\vrule height11pt width0pt
 3 & 4.370\,413\,478\,422(3)  & 4.370\,420\,247\,640(2)  \\
   & 4.370\,413\,470(7)$^d$   & 4.370\,420\,247(2)$^c$ \\
 4 & 4.370\,417\,339\,045(4)  & 4.370\,421\,238\,038(4) & 4.370\,421\,511\,306(3) & 4.370\,421\,527\,144(3) \\
 5 & 4.370\,419\,597\,74(2)   & 4.370\,421\,809\,90(2)   \\
\hline\hline
\end{tabular}
\vspace{-1mm}
\begin{flushleft}
\hspace{19mm}$^{a}$Drake and Goldman \cite{DrakeCJP}\\
\hspace{19mm}$^{b}$Yerokhin and Pachucki \cite{Pachucki10}\\
\hspace{19mm}$^{c}$Korobov \cite{Kor12} \\
\hspace{19mm}$^{d}$Wienczek \emph{et al.} \cite{Pachucki19}
\end{flushleft}
\end{center}
\end{table}

Typically $[A_1,A_2] = [3,6]$, and $n_{\rm max} = 5\!-\!7$, that corresponds to the photon energy interval $k\in[0,10^4]$.
\item For other pairs of $(i,j)$ we take the similar basis sets as in 2. It is worthy to note that for identical particles this step should be omitted, since, as it follows from Eq.~(\ref{eq:psi1}), there is no singular behaviour of $\psi_1(r_{ij},\cdot)$ for small $r_{ij}$.
\end{enumerate}

After the complete set of basis functions is constructed, we diagonalize the matrix of the Hamiltonian $H_I$ for intermediate states to get a set of (pseudo)state energies, $E_m$, and then to calculate $\left\langle 0 |i\mathbf{J}| m \right\rangle$. These two sets of data are enough to restore $J(k)$:
\begin{equation}
J(k) = -\sum_m
     \frac{\left\langle 0| i\mathbf{J} |m\right\rangle^2}{E_0\!-\!E_m\!-\!k}\>,
\end{equation}
and to integrate the low energy part of the numerator $\mathcal{N}(L,v)$
\begin{equation}
\int_0^{E_h}\,k\,dk
   \left\langle
      \mathbf{J}\left(\frac{1}{E_0\!-\!H\!-\!k}+\frac{1}{k}\right)\mathbf{J}
   \right\rangle
   +\int_{E_h}^\Lambda\,\frac{dk}{k}\,
   \left\langle
      \mathbf{J}\frac{(E_0\!-\!H)^2}{E_0\!-\!H\!-\!k}\mathbf{J}
   \right\rangle.
\end{equation}

From thus obtained $J(k)$ we extrapolate coefficients of asymptotic expansion
\begin{equation}\label{asy_num}
f_{\rm fit}(k)=\sum_{m=1}^M \frac{C_{1m}\sqrt{k}\!+\!C_{2m}\ln{k}\!+\!C_{3m}}{k^{m+3}},
\end{equation}
which is taken in the same form as in analytic expansion for the hydrogen atom. The leading order terms of $J(k)$ are obtained from Eq. (\ref{expansion}). That allows to calculate the high energy part of the numerator
\[
\int_\Lambda^\infty\,\frac{dk}{k}\,
   \left\langle
      \mathbf{J}\frac{(E_0\!-\!H)^2}{E_0\!-\!H\!-\!k}\mathbf{J}
   \right\rangle\>.
\]

\section{Results}

The results of our numerical calculations are presented in Table \ref{tb:results}, where the infinite mass of the nucleus is assumed. As is seen from the data our results are in a good agreement with previous calculations \cite{DrakeCJP,Pachucki10}, but exceed them in numerical precision. In order to calculate data for comprehensive analysis of the leading order radiative corrections we have to consider mass dependence of the Bethe logarithm and extrapolation to high $n$ states. That will be done in the subsections below.

\begin{table}[b]
\caption{Mass dependence of the Bethe logarithm. Coefficients of expansion (\ref{eq:mass})}\label{tab:mass}
\begin{center}
\begin{tabular}{c@{\hspace{8mm}}r@{\hspace{8mm}}r}
\hline\hline
\vrule height9pt width0pt
state & $a_1$~~~~~~~~~ & $a_2$~~~~~~~~ \\
\hline
\vrule height11pt width0pt
 $1^1S$  &  $ 0.9438944[-01]$ &  $-0.16501[+00]$  \\
 $2^1S$  &  $ 0.1773442[-01]$ &  $-0.34889[-01]$  \\
 $2^3S$  &  $ 0.4785558[-02]$ &  $-0.95888[-02]$  \\
 $2^1P$  &  $-0.3553442[-02]$ &  $ 0.95390[-02]$  \\
 $2^3P$  &  $ 0.8709662[-02]$ &  $ 0.37926[-02]$  \\
 $3^1S$  &  $ 0.5386215[-02]$ &  $-0.10813[-01]$  \\
 $3^3S$  &  $ 0.1071013[-02]$ &  $-0.29114[-02]$  \\
 $3^1P$  &  $-0.1004733[-02]$ &  $ 0.20368[-02]$  \\
 $3^3P$  &  $ 0.2536963[-02]$ &  $-0.11556[-02]$  \\
 $3^1D$  &  $-0.6358710[-05]$ &  $ 0.36083[-03]$  \\
 $3^3D$  &  $ 0.3792498[-04]$ &  $-0.22445[-03]$  \\
 $4^1S$  &  $ 0.2264400[-02]$ &  $-0.45874[-02]$  \\
 $4^3S$  &  $ 0.3763586[-03]$ &  $-0.12354[-02]$  \\
 $4^1P$  &  $-0.4158826[-03]$ &  $ 0.78650[-03]$  \\
 $4^3P$  &  $ 0.1074222[-02]$ &  $-0.72991[-03]$  \\
 $4^1D$  &  $-0.6179997[-05]$ &  $ 0.17459[-03]$  \\
 $4^3D$  &  $ 0.1964547[-04]$ &  $-0.17124[-03]$  \\
 $4^1F$  &  $ 0.3926256[-05]$ &  $-0.11805[-03]$  \\
 $4^3F$  &  $ 0.3612599[-05]$ &  $-0.11625[-04]$  \\
 $5^1S$  &  $ 0.1151476[-02]$ &  $-0.23793[-02]$  \\
 $5^3S$  &  $ 0.1710350[-03]$ &  $-0.51711[-03]$  \\
 $5^1P$  &  $-0.2113686[-03]$ &  $ 0.38744[-03]$  \\
 $5^3P$  &  $ 0.5516224[-03]$ &  $-0.41064[-03]$  \\
 $5^1D$  &  $-0.4174068[-05]$ &  $ 0.75987[-03]$  \\
 $5^3D$  &  $ 0.1077559[-04]$ &  $-0.64813[-04]$  \\
 $6^1S$  &  $ 0.6621675[-03]$ &  $-0.23261[-02]$  \\
 $6^3S$  &  $ 0.9116608[-04]$ &  $-0.64837[-04]$  \\
 $6^1P$  &  $-0.1225350[-03]$ &  $-0.43729[-03]$  \\
 $6^3P$  &  $ 0.3206520[-03]$ &  $-0.33082[-03]$  \\
\hline\hline
\end{tabular}
\end{center}
\end{table}

\subsection{Finite mass}

For the case of a finite mass $M$ of the nucleus, the results of Table \ref{tb:results} should be somehow modified. The most simple way is to use an expansion
\begin{equation}\label{eq:mass}
\beta_M = \beta_{\infty}+\ln{(\mu/m_e)}+a_1(m_e/M)+a_2(m_e/M)^2+\cdots
\end{equation}
The major contribution, $\ln{(\mu/m_e)}$, comes from the scaling of the electron wave function due to the finite mass effect, where $\mu\!=\!m_eM/(M\!+\!m_e)$ is the reduced mass of an electron. The next linear term may be calculated using the first-order perturbation of the Bethe logarithm by the mass-polarization operator, see \cite{Pachucki10}. In our case we utilized a more trivial approach. Since our code was written for a general case of three particles of finite masses, we performed calculations of the Bethe logarithm for two consecutive points of the variable $M$, namely, for $M_1=M_{^4\!\rm He}=7294.29954136\,m_e$ and $M_2=M_1/2$. For a mass of the alpha particle we take the CODATA14 recommended value \cite{CODATA14}. The two coefficients of the linear and quadratic terms in (\ref{eq:mass}) have been obtained from a set of two linear equations. The linear coefficients are in a very good agreement with the corresponding coefficients calculated by Yerokhin and Pachucki in \cite{Pachucki10}, as may be expected. We also checked the quality of our approximation for the $1^{1\!}S$ state using the least squares approximation for $a_1$ and $a_2$ from the three point data with the third data point $M_3=M_1/3$. The least squares result: $a_1 = 0.09438942$ and $a_2 = -0.1649032$, agrees well with the simple approximation and demonstrates that the data presented in Table \ref{tab:mass} is sufficient for precise determination of the Bethe logarithm for real cases of the helium atom for various nuclear isotopes.

\subsection{Asymptotic expansions for Rydberg states}

For the Rydberg states the Bethe logarithm may be calculated using the asymptotic expansion \cite{Drake92,Drake01}
\begin{equation}\label{asy1}
\beta(1,s;n,l;\,^{1,3\!}L) = \beta_{1s} +\ln{Z^2} + \left(\frac{Z-1}{Z}\right)^4\frac{\beta_{nl}}{n^3}
   +\frac{0.316205(6)}{Z^6}\left\langle x^{-4} \right\rangle + \Delta\beta(1,s;n,l;\,^{1,3\!}L),
\end{equation}
where $x$ is the distance between the outer Rydberg electron and nucleus. In the approximation of an electron in a field of the effective charge $Z^*=1$ this quantity given by expression Eq.~(\ref{asy1}) is valid for states with $L>0$ \cite{Drake92}. Here $\beta_{1s}=2.984128555765$ is the Bethe logarithm for the ground state of a hydrogen atom, and $\Delta\beta(1,s;n,l\,^{1,3}L)$ takes into account contributions from the higher multipole moments \cite{Drake01} (see Eqs.~(19)-(20)).

In our case we use a slightly different expansion
\begin{equation}\label{asy2}
\beta(1,s;n,l) = \beta_{1s} +\ln{Z^2} + \frac{1}{n^3}\sum_{j=0}^2\frac{c_j}{n^j}
\end{equation}
where the coefficients $c_i$ are obtained by the method of least squares \cite{Lowson} and data from Table \ref{tb:results}. To get a proper extrapolation in case of $S$ states we have used data points for states $n=3$ and up to $n=7$ in order to eliminate the nonmonotonic behaviour of the $\beta(n)$ function for the first two states. Coefficients of the asymptotic expansion (\ref{asy2}) are presented in Table \ref{tab:asy}. We estimate that the values obtained using expansion (\ref{asy2}) for the Bethe logarithm of the Rydberg states should be accurate up to 7-8 digits.

For the states with $L\ge3$ the asymptotic formula (\ref{asy1}) works well. Say, the results for the $4F$ states: $\beta(4^1F)=4.370421511(13)$ and $\beta(4^3F)=4.370421527(15)$, agree to the 10-figure accuracy with our results presented in Table \ref{tb:results}. For further discussions we refer readers to \cite{Drake01}.

\begin{table}[t]
\caption{The Bethe logarithm for the Rydberg states. Coefficients of the asymptotic expansion (\ref{asy2}).}\label{tab:asy}
\begin{center}
\begin{tabular}{r@{\hspace{8mm}}r@{\hspace{5mm}}r@{\hspace{5mm}}r@{\hspace{5mm}}r@{\hspace{5mm}}r@{\hspace{5mm}}r}
\hline\hline
 & $n^1S$~~~ & $n^3S$~~~ & $n^1P$~~~ & $n^3P$~~~ & $n^1D$~~~~ & $n^3D$~~~~ \\
\hline
$c_0$ & $-$0.03151 & $-$0.03530 & $-$0.004857 & $-$0.006620 & $-$0.0006212 & $-$0.0003348 \\
$c_1$ & $-$0.01874 & $-$0.04245 &    0.003677 &    0.001407 &    0.0009306 &    0.0012734 \\
$c_2$ &    0.03409 &    0.01840 &    0.001809 &    0.009856 &    0.0005059 & $-$0.0014552 \\
\hline\hline
\end{tabular}
\end{center}
\end{table}

\subsection{Conclusions}

In conclusion, we want to summarize the main results of our work. First, the new more accurate values of the Bethe logarithm have been obtained numerically for a wide range of states with the precision which exceeds the published data by at least three orders of magnitude. These data allowed us to make asymptotic extrapolation to the states with higher $n$, a principal quantum number of the excited electron. Along with results obtained by Drake in \cite{Drake01}, our data cover the whole set of bound states in a helium atom.

\section*{Acknowledgements} The work has been carried out under financial support of the Russian Foundation for Basic Research Grant No.~19-02-00058-a, author also acknowledges support of the "RUDN University Program 5-100".

\end{document}